\documentclass{article}
\usepackage{neurips_2023}
\usepackage{subcaption}
\usepackage{graphicx}
\usepackage{float}
\usepackage[utf8]{inputenc}
\usepackage[T1]{fontenc}    
\usepackage{hyperref}       
\usepackage{url}           
\usepackage{booktabs}     
\usepackage{amsfonts}    
\usepackage{nicefrac}    
\usepackage{microtype}   
\usepackage{xcolor}         
\title{A Decadal Analysis of the Lead-Lag Effect in the NYSE}
\begin{document}
\maketitle
\section{Introduction}
\large
As is widely known, the stock market is a complex system in which a multitude of factors influence the performance of individual stocks and the market as a whole. One method for comprehending—and potentially predicting—stock market behavior is through network analysis, which can offer insights into the relationships between stocks and the overall market structure. In this paper, we seek to address the question: Can network analysis of the stock market, specifically in observation of the lead-lag effect, provide valuable insights for investors and market analysts? This inquiry is both interesting and pertinent for several reasons.

Firstly, grasping the relationships between stocks and the overall market structure can aid investors in making more informed, and potentially more profitable, decisions regarding their investments. Additionally, network analysis may offer new tools for monitoring the stock market and identifying trends or potential risks. Through this investors will be able to observe the bad returns of a leading stock and therefore draw conclusions about the effects this will have on other closely tied securities.

To tackle this question, we will build upon two previous studies—of which both observed a power-law distribution in stock returns. The first, performed by researchers at HIT China, constructed a network for the US stock market based on a model in which two stocks were said to be connected if their returns fell in the respective returns of each other over some period to some threshold of accuracy. The study found that through the use of this technique, they were able to select a reasonably performing portfolio that beat the S\&P 500 [1]. The second study, performed by a group of researchers at Iran's Tehran University, employed community detection techniques to construct a correlation network. They discovered that the resulting communities were consistent with market sectors classified using the Standard Industrial Classification code. They also utilized network analysis to construct visualizations of the return correlations among various public stocks, which offered an intuitive way to examine the overall correlation structure of different public stocks and identify key market segments [2].

While these prior works provide valuable insights into the network structure of the stock market, there remains much to expand on this approach. In our research, we concentrate on the lead-lag effect, which refers to the phenomenon where the returns of one stock lead, over some period---referred to as the "lag"---the returns of another stock. By analyzing the lead-lag effect within the network of the stock market, we aim to offer insights into the dynamics of market behavior and inform investment strategies.
\section{Methodology}
We realize that the lead-lag relationship is a "strategy-enhancer," but not a strategy in itself. Thus, we combined it with a known, semi-reliable, fundamental strategy: the  Capital Asset Pricing Model (CAPM) as well as with another network theory technique, out-degree. Simply put, CAPM is an effective way to compute the expected return of a given security based on the market risk premium and the systematic risk of the security and out-degree in a measure of the number of edges leaving a given node. With out-degree, specifically, we looked at the out-degree of the lagger---assuming that a lagger with a lower out-degree will receive greater input from its few leaders. CAPM is normalized around the market rate, so we curve our out-degree scores and then weigh the two methods 70-30, CAPM and out-degree respectively. From this, we choose the top six stock pairs that have the highest absolute sum, these are the stocks that we will trade over the quarter. With the coupling of these two strategies, we selected a portfolio and compared its returns to the S\&P 500.
\subsection{Data Collection and Preprocessing}
We first constructed a network of the market using the stock return data of the S\&P 500. This involved building a script that web scraped for relevant data. Using clever list manipulation and Python's  {\fontfamily{qcr}\selectfont
numba} package, we were able to tame the large computational time that comes with processing the data of 500 stocks for ten years. Correspondingly, there are roughly 250 trading days per year so making sure that our data cubes matched imported S$\&$P 500 time series data was critical to ensure a reliable comparison.

The data used in subsequent steps is sourced from a collections of files we generated. One contains daily stock prices for all assets in the S$\&$P 500, another contains the lead-lag preprocessed adjacency matrices in the form of a tensor, and another contains quarterly CAPM data for all assets.
\subsection{Identifying Lead-Lag Pairs}
We then created our algorithm, selecting connection thresholds found in the literature, first by constructing a 500x500 matrix of the closing prices of the stocks. If, over consecutive time periods, the return of stock {\fontfamily{qcr}\selectfont
i} stays within a given interval of stock {\fontfamily{qcr}\selectfont
j} then we consider stock {\fontfamily{qcr}\selectfont
j} to lead stock {\fontfamily{qcr}\selectfont
i}, where the time period is called the lag. If a stock leads another stock we placed a {\fontfamily{qcr}\selectfont
1} in our adjacency matrix \footnote{Which is a pseudo adjacency matrix, as although it resembles an adjacency matrix it is not symmetric nor does it have {\fontfamily{qcr}\selectfont
1}'s completely along the diagonal.}. Noticeably, stocks tend to lead themselves, but because of volatility, certain stocks do not. 

To identify the leader-lagger pairs, a 2D array is created from the summed data, with the diagonal masked to ignore self-pairs, in other words, we ignore stocks leading or lagging themselves. The top 20 largest values in the masked array are then extracted, and the corresponding row and column indices are identified. These indices are used to create a dictionary of leader-lagger pairs, where the leader is the stock that leads in price movement, and the lagger is the stock that follows. Using CAPM information of laggers, we update this dictionary quarterly to select the top six best candidates\footnote{These pairs have laggers with the lowest out-degree as mentioned in the Introduction.} according to our strategy and trade with them through the quarter. 
\subsection{Trading Strategy and Portfolio Simulation}
The trading strategy used in this study is based on the following rules:
\begin{itemize}
    \item If a leader stock's price increases by at least the buy threshold from the previous day, buy the corresponding lagger stock.
    \item If a leader stock's price decreases below a trailing stop level, sell the corresponding lagger stock.
\end{itemize}
The study assumes an initial investment of 500,000 USD, with the investment amount evenly distributed among the lagger stocks. A portfolio dictionary is initialized to track the number of shares held for each lagger stock. For each trading day, the strategy's rules are applied, and the portfolio is updated quarterly with new CAPM data. Buying threshold and trailing stop percentage are used in subsequent parameter exploration and used to maximize portfolio return.
\subsection{Performance Evaluation}
The performance of the trading strategy is evaluated by calculating the portfolio return over the period. The return is calculated as the difference between the final value and the initial value of the portfolio, divided by the initial value. The portfolio return is compared with the S$\&$P 500 index return over the same period to provide context for the strategy's performance using the {\fontfamily{qcr}\selectfont
yfinance} API. To visualize the performance, a line chart is created, showing the cumulative returns for both the portfolio and the S$\&$P 500 index, see Fig. \ref{fig:portfolio}
\subsection{Visualization of Leader-Lagger Relationships}
To visualize the relationships between leader and lagger stocks, a directed graph is created using Python's {\fontfamily{qcr}\selectfont
networkx} package, see Fig. \ref{fig:networks}. The nodes in the graph represent stocks, while the directed edges represent the leader-lagger relationships. Leader stocks are colored red, and lagger stocks are colored blue. The directed graph is displayed using using Python's {\fontfamily{qcr}\selectfont
matplotlib} package.
\section{Results}
Our created portfolio beat the S$\&$P 500 by about ten percent over the course of the last four fiscal quarters, ubiquitously known to be a bear market, see Fig. \ref{fig:bearPlot}.\footnote{We analyzed the lead-lag effect on data spanning from March 15, 2022 to March 15, 2023.} Moreover, during a bull market, such as during the second and third fiscal quarters of 2021, back-testing delivered a larger positive portfolio return when compared to the S$\&$P 500 during the same period, see Fig. \ref{fig:bullPlot}. As well, in the major bull market of 2021 we beat the S$\&$P 500 by about ten percent once again.
\subsection{Network and Graph Model}
\begin{figure}[H]
    \centering
    \begin{subfigure}{0.45\textwidth}
        \centering
        \includegraphics[width=\linewidth]{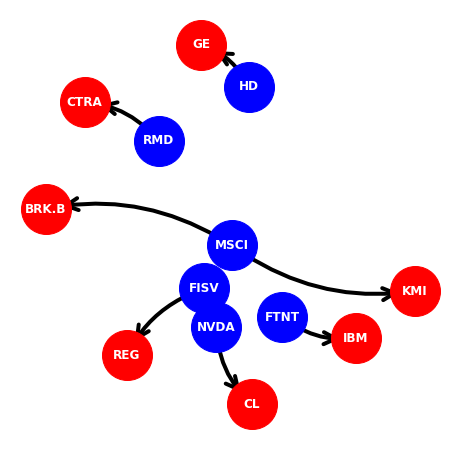}
        \caption{Bull market lead-lag pairs.}
        \label{fig:bullNetwork}
    \end{subfigure}
    \hfill
    \begin{subfigure}{0.45\textwidth}
        \centering
        \includegraphics[width=\linewidth]{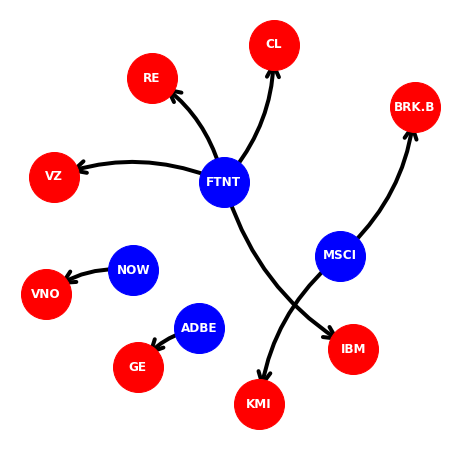}
        \caption{Bear market lead-lag pairs.}
        \label{fig:bearNetwork}
    \end{subfigure}
    \caption{Network graphs of lead-lag pairs for the first day of trading. Blue nodes represent laggers with directed edges connecting them to red nodes representing leaders.}
    \label{fig:networks}
\end{figure}

Evident in Fig. \ref{fig:bearNetwork}, a network plot of our lead-lag pairs on the first day of trading\footnote{It would be interesting to study the temporal evolution of these plots, specifically the change in unique lead-lag pairs.} during a bear market, multiple leaders led the same lagger and the strongest signals often came from one lagger. Interestingly enough, lead-lag pairs did not always form along industry lines, in contradiction with previous research.\footnote{Likely because we used a shorter lag period.} Correspondingly, we observe a greater number of lead-lag pairs during the first trading day of the bull market, see Fig. \ref{fig:bullNetwork}. Since our analysis is centered around a directed graph, we are precluded from studying the assortment of network centrality measures, namely, eigenvector centrality.\footnote{In future work, the adjacency matrix can be made symmetric and these measures can be deployed.}
\subsection{Portfolio Simulation}
\begin{figure}[H]
    \centering
    \begin{subfigure}{0.495\textwidth}
        \centering
        \includegraphics[width=\linewidth]{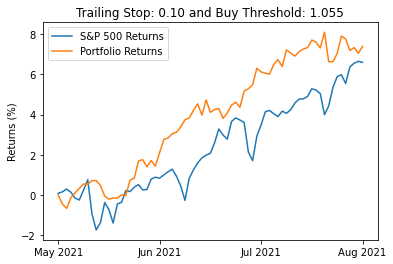}
        \caption{Bull market of Q2 and Q3 of 2021.}
        \label{fig:bullPlot}
    \end{subfigure}
    \hfill
    \begin{subfigure}{0.495\textwidth}
        \centering
        \includegraphics[width=\linewidth]{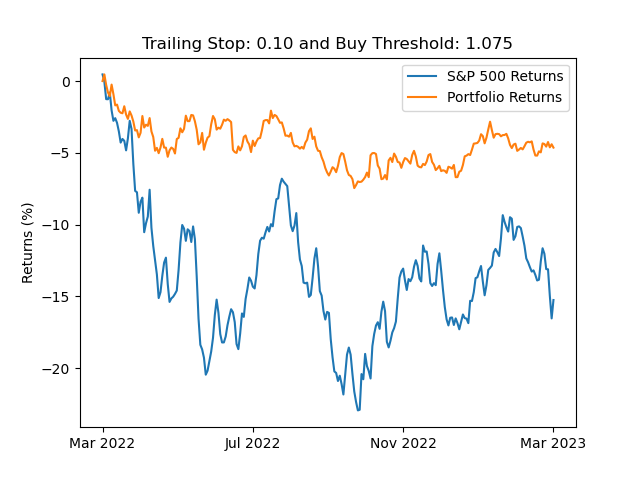}
        \caption{Bear market of the past four quarters.}
        \label{fig:bearPlot}
    \end{subfigure}
    \caption{Back-testing of portfolio returns compared to historical returns of the S$\&$P 500  ($^\wedge$GSPC).}
    \label{fig:portfolio}
\end{figure}

Over the course of the last four fiscal quarters and during the summer of 2021, our portfolio returns mirrored the ebb and flow of the S$\&$P 500's historical returns, see Fig. \ref{fig:portfolio}. Since bear markets are hard to predict it is was reasonable for our portfolio simulation to not experience positive returns\footnote{It is quite difficult without use of leverage to experience positive gains in a bear market and our portfolio does not include an options trading functionality.}, though we lose roughly five percent in value while the S$\&$P 500 approximately loses fifteen percent, see Fig. \ref{fig:bearPlot}.\footnote{Hence, someone who did not invest at all would have kept the value of their capital while our portfolio and the S$\&$P 500's value diminished, not accounting for inflation.} Moreover, during a bull market, we observe a positive return roughly twelve percent greater than the S$\&$P 500 during the same period, see Fig. \ref{fig:bullPlot}.
\subsection{Parameter Exploration}
Naturally, we were prompted to conduct basic parameter exploration with the hopes of maximizing our return given our risk preferences amongst other variables.

\subsubsection{Bear Market}
We observed that when the buy threshold was set to a value greater than or equal to 1.13, the model never engaged in any trades, thereby resulting in no portfolio returns. This suggests that a buy threshold of 1.13 or higher is too conservative for the model to capitalize on market opportunities.

\begin{figure}[H]
    \centering
    \begin{subfigure}{0.495\textwidth}
        \centering
        \includegraphics[width=\linewidth]{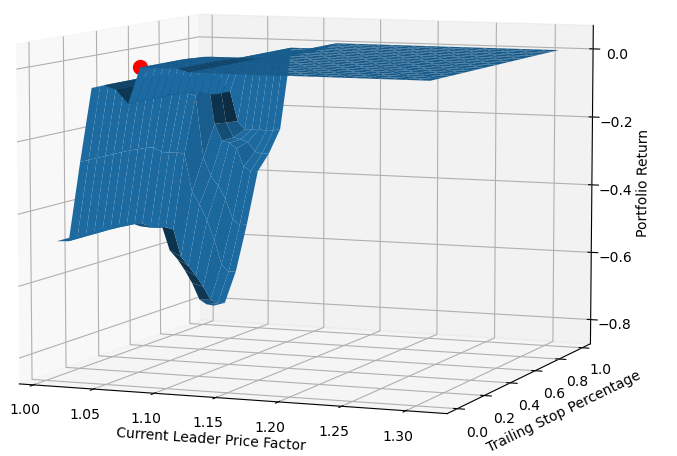}
        \caption{Contour plot for bear market parameters.}
        \label{fig:contourPlot}
    \end{subfigure}
    \hfill
    \begin{subfigure}{0.495\textwidth}
        \centering
        \includegraphics[width=\linewidth]{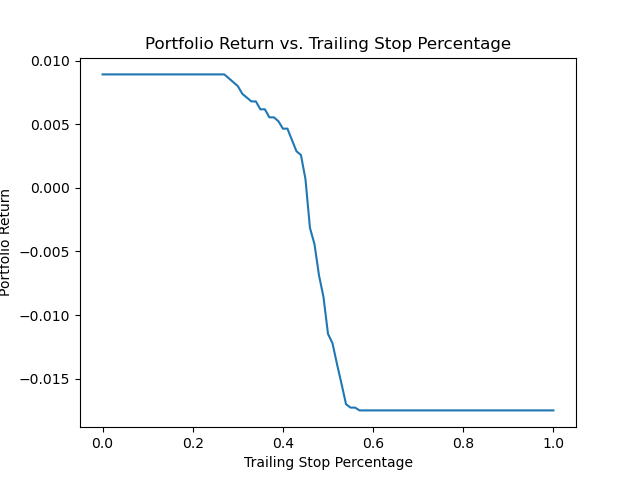}
        \caption{2D cross-section of bear contour plot, holding buy threshold constant.}
        \label{fig:buyPlot}
    \end{subfigure}
    \label{fig:parameters}
    \begin{subfigure}{0.495\textwidth}
        \centering
        \includegraphics[width=\linewidth]{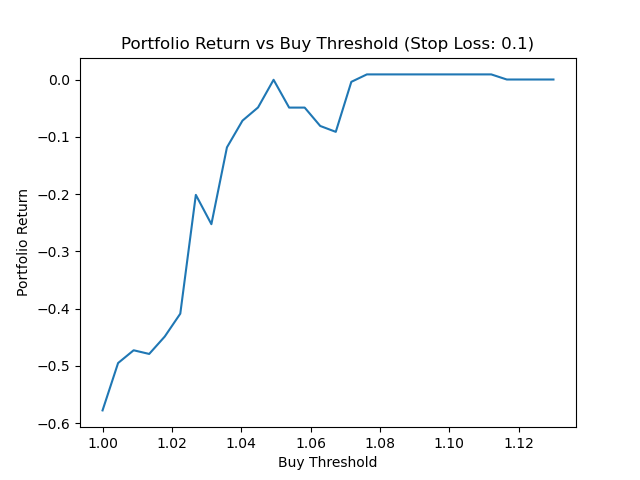}
        \caption{2D cross-section of bear contour plot, holding stop loss threshold constant.}
        \label{fig:stopPlot}
    \end{subfigure}
    \caption{Mesh grid plot of trailing stop percentage vs. buy threshold vs. portfolio returns in a bear market.}

\end{figure}

Furthermore, we found that the highest portfolio value was attained when the trailing stop percentage was set to a value less than or equal to 0.40. At this threshold, all trades were sold on a daily basis, indicating a high turnover strategy. This result is demonstrated in the contour plot presented in Fig. \ref{fig:contourPlot}.

In contrast, when the buy threshold was set to a value less than or equal to 1.07, the model traded more frequently, exposing the portfolio to increased risk. The 2D cross-section of the contour plot in Fig. \ref{fig:stopPlot} shows the impact of the buy threshold on trading frequency and risk exposure. As a result we used a trailing stop percentage of 0.10 in our portfolio simulations and toggled the buying threshold according to whether or not we were trading in a bear or bull market, see Fig. \ref{fig:portfolio}.

\subsubsection{Bull Market}

The behavior observed in the bull market was, as expected, different to that observed in a bear market.

\begin{figure}[H]
    \centering
    \begin{subfigure}{0.495\textwidth}
        \centering
        \includegraphics[width=\linewidth]{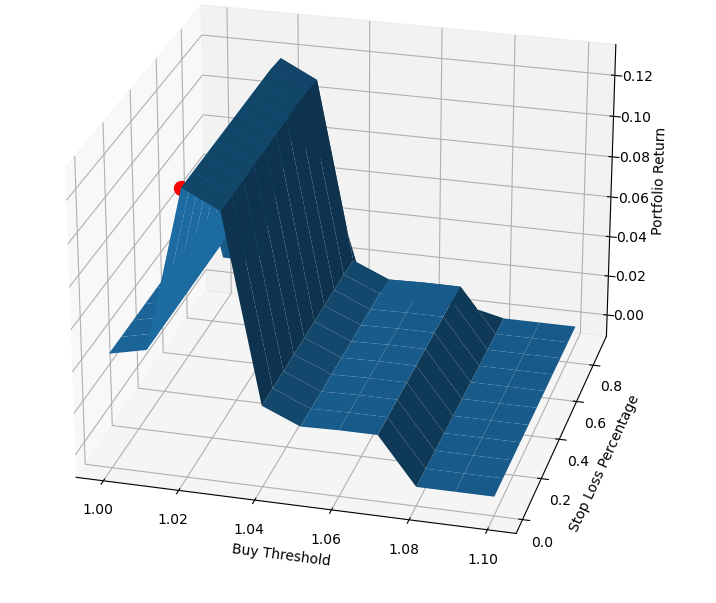}
        \caption{Contour plot for bull market parameters.}
        \label{fig:bullcontourPlot}
    \end{subfigure}
    \hfill
    \begin{subfigure}{0.495\textwidth}
        \centering
        \includegraphics[width=\linewidth]{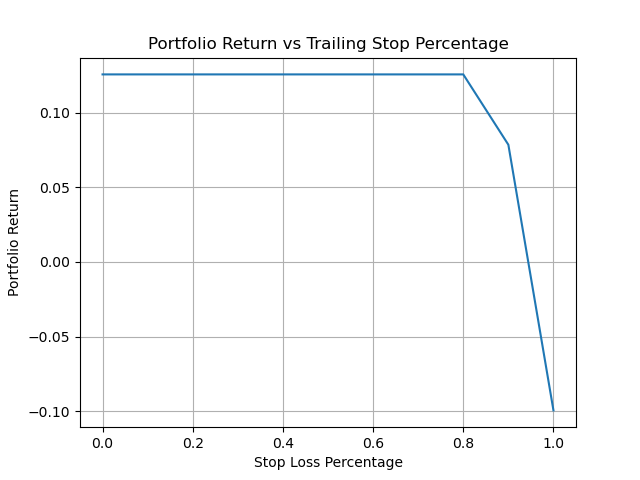}
        \caption{2D cross-section of bull contour plot, holding buy threshold constant.}
        \label{fig:bullbuyPlot}
    \end{subfigure}
    \label{fig:buyparameters}
    \begin{subfigure}{0.495\textwidth}
        \centering
        \includegraphics[width=\linewidth]{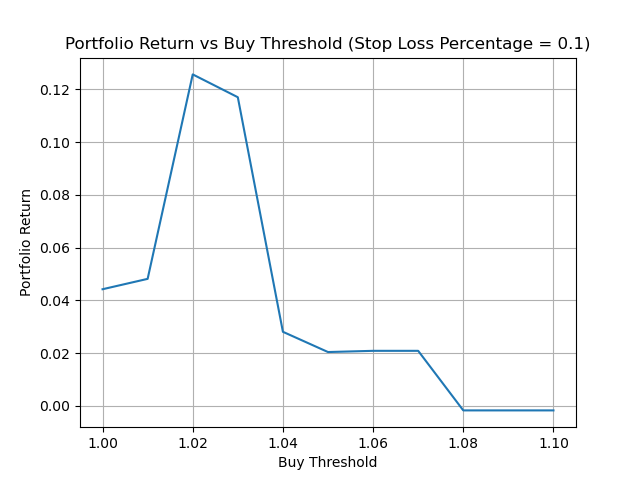}
        \caption{2D cross-section of bull contour plot, holding stop loss threshold constant.}
        \label{fig:bullstopPlot}
    \end{subfigure}
    \caption{Mesh grid plot of trailing stop percentage vs. buy threshold vs. portfolio returns in a bull market.}
\end{figure}

We observed that when the buy threshold was set to a value greater than or equal to 1.08\footnote{Although this seems contrary to intuition this is an artifact of stocks in 2021 having lesser daily gains but but greater gains overall. Moreover, 2022 had more visible return volatility but overall maintained this behavior.}, the model refrained from engaging in any trades (Fig \ref{fig:bullstopPlot}), consequently yielding no portfolio returns. This phenomenon may be attributed to the fact that in a bull market, the model perceives a buy threshold of 1.08 or higher as overly conservative, thus missing out on potential market opportunities. Additionally, we discovered that the trailing stop percentage negatively impacted the portfolio when set above 0.80 (Fig \ref{fig:bullbuyPlot}). This could be due to the fact that a higher trailing stop in a bull market may lead to premature exits from profitable positions, limiting the potential gains.

Interestingly, our study found that the optimal buy threshold was set at 1.02, where the return increased from 1 to 1.02 and subsequently decreased up to 1.08( Fig \ref{fig:bullcontourPlot}). This suggests that a buy threshold set within this range allows the model to effectively capitalize on the upward market trend while minimizing the risk of overexposure. The reason for this increase in return up to 1.02 and subsequent decrease can be attributed to the balance between capturing gains during the bull market and avoiding excessive risk exposure. At a buy threshold of 1.02, the model is able to participate in the market's upward momentum, while still maintaining a relatively conservative approach to minimize potential losses. As the buy threshold increases beyond 1.02, the model becomes increasingly conservative, leading to missed opportunities and a decline in returns.
In both bear and bull markets, our findings indicate that an optimal trading strategy would involve setting the buy threshold and trailing stop percentage within specific ranges to balance the trade-offs between portfolio returns, trading frequency, and risk exposure. Further research is warranted to refine these ranges and explore additional parameters that may influence the performance of the model.
\section{Analysis and Considerations}
In our study on the lead-lag effect, there were several considerations that we took into account
 while developing our model. Firstly, the lead-lag threshold that we decided upon was not extensively tested, 
  which could potentially affect the robustness of our findings. 
  There is also the issue of overfitting when determining the buy threshold and stop loss parameters. Through parameter estimation, we identified values that appeared to be profitable in both bull and bear markets. However, it is crucial to exercise caution when generalizing these findings to a larger sample, as overfitting can lead to overly optimistic performance estimates that may not hold true in different market conditions. 
  Additionally, our chosen fundamental strategy, the Capital Asset Pricing Model (CAPM),
  is based on certain unrealistic assumptions that may not accurately reflect real-world trading conditions. Lastly, it is important to note that due to the frequent trading nature of our model, 
  commission costs must be considered as they could significantly impact the overall performance of the model. These considerations highlight the challenges and limitations
   faced when attempting to develop an effective trading model that accurately captures the complexities of the lead-lag effect.

\section{Conclusions}
Based on our analysis of the lead-lag effect in the NYSE, we draw several significant conclusions. Firstly, our model, which combines network analysis with fundamental strategies, can potentially outperform the market in predicting stock performance. However, we acknowledge that our results are based on a specific subset of years and may not be able to be generalized. 

Secondly, our finding that lead-lag pairs often form outside of industry lines contradicts previous literature and suggests that investors may benefit from examining unconventional connections. 

Thirdly, we demonstrate that our strategy can be customized to fit individual investors' risk preferences by adjusting the buy threshold. In bearish markets, investors can raise the threshold to experience the market's ups and downs to a lesser degree, while in growth-favoring markets, investors can lower the threshold to expose themselves to more risk and potential return. 

Overall, our study highlights the usefulness of network analysis in predicting stock performance and the importance of considering unconventional connections in investment decisions. Furthermore, our approach allows for customization to meet individual risk preferences, demonstrating its practical relevance to investors.

\section{Future Work}
There are several potential areas for future work and unanswered questions that could be addressed in further research. First, our study did not thoroughly explore the impact of different time periods on the effectiveness of the lead-lag effect. Investigating whether the relationships observed in our study hold true over longer time periods or during different economic cycles and market conditions would be an interesting avenue for future research.

In addition, our study focused on the use of network analysis combined with fundamental strategies in predicting stock performance. However, there are several other strategies and factors that could be incorporated into the analysis. For instance, investigating the effectiveness of incorporating technical analysis, sentiment analysis, or macroeconomic indicators in combination with network analysis could yield interesting results. 

Moreover, it would be more realistic to incorporate weighted edges and a distributed value for the lead-lag coefficient in our model, taking into account the differences in market capitalization among stocks. Specifically, assigning higher numbers in the matrix to stocks that were extremely lagged over a given period, and assigning less weight to the edges of stocks with higher market capitalization. As well, it would be helpful to investigate the effectiveness of different lead-lag thresholds in predicting stock performance.

\section*{References}
[1] Li, Y., Wang, T., Sun, B., \& Liu. Detecting the lead–lag effect in stock markets: Definition, patterns, and investment strategies. Financial Innovation, 2022.

[2] A. Namaki, A.H. Shirazi, R. Raei, and G.R. Jafari. Network analysis of a financial market based on genuine correlation and threshold method. Physica A: Statistical
Mechanics and its Applications, 2011.

[3] R. D. Smith. The Spread of the Credit Crisis: View
from a Stock Correlation Network. Journal of Korean
Physical Society, 54:2460, June 2009.

\medskip
\small
\end{document}